\documentclass[10pt,english,prb,amsmath,amssymb,prb,showkeys,superscriptaddress,twocolumn,showpacs,floatfix]{revtex4-2}
\usepackage{float}
\usepackage{graphicx}
\usepackage{tabularx}
\usepackage{epstopdf}
\usepackage{dcolumn}
\usepackage[colorlinks=true,linkcolor=blue ,citecolor=blue,urlcolor=blue]{hyperref}
\usepackage{multirow}
\usepackage[usenames,dvipsnames]{xcolor}
\usepackage{soul}
\usepackage{braket}
\usepackage{bm}
\usepackage{nicefrac}
\usepackage{siunitx}
\usepackage{color,transparent}
\usepackage[utf8]{inputenc}
\usepackage{upgreek}

\usepackage[version=4]{mhchem}
\usepackage{chemformula}
\usepackage{amsmath}   
\usepackage{mathtools} 
\usepackage{amssymb}   
\usepackage{mathrsfs}  
\usepackage{textcomp}  
\usepackage{accents}   
\usepackage{bm}        
\usepackage{soul}
\usepackage{relsize}
\usepackage{graphicx}  
\usepackage{epstopdf}  
\usepackage{subfigure} 
\usepackage{makecell}
\usepackage{hhline}
\usepackage{cellspace}
\usepackage{array}     
\usepackage{tabularx}  
\usepackage{multirow}  
\usepackage{booktabs}  
\usepackage{dcolumn}   
\usepackage{gensymb}   
\usepackage{setspace}
\usepackage{scrextend}
\usepackage{float}

\begin{document}
\setlength{\heavyrulewidth}{0.08em}
\setlength{\lightrulewidth}{0.05em}
\setlength{\cmidrulewidth}{0.03em}
\setlength{\belowrulesep}{0.65ex}
\setlength{\belowbottomsep}{0.00pt}
\setlength{\aboverulesep}{0.40ex}
\setlength{\abovetopsep}{0.00pt}
\setlength{\cmidrulesep}{\doublerulesep}
\setlength{\cmidrulekern}{0.50em}
\setlength{\defaultaddspace}{0.50em}
\setlength{\tabcolsep}{4pt}
\title{First-principles investigation of small polarons in rhombohedral \ce{NaNbO3}}
\author{Mohammad Amirabbasi}
\email{amirabbasi@mm.tu-darmstadt.de}

\author{Lorenzo Villa}
\author{Elaheh Ghorbani}
\author{Jochen Rohrer}
\author{Karsten Albe}
\email{albe@mm.tu-darmstadt.de}

\address{Technical University of Darmstadt, Institute of Materials Science, Materials Modelling Division, Otto-Berndt-Straße 3, Darmstadt D-64287, Germany}

\begin{abstract}
Sodium niobate (\ce{NaNbO3}) is a perovskite oxide and a key component of emerging lead-free antiferroelectric capacitors for high-energy-density applications. However, its performance can be hindered by irreversible phase transitions and leakage currents associated with low electrical resistivity.
Defect and doping engineering offers a potential way to overcome these problems, but its use requires a detailed understanding of electronic, ionic, and polaron charge-compensation mechanisms, where the role of polarons remains largely unexplored.
Here, we investigate the stability of small hole and electron polarons in rhombohedral \ce{NaNbO3}, which is a structurally well-defined model system that avoids lattice-dynamical instabilities. Trapping energies are calculated using density-functional theory corrected by a Hubbard~$U$, using the enforced-piecewise-linearity approach including finite-size scaling.
For the small hole-polaron centered on O-2$p$ orbital, we find a trapping energy of $-$0.65~(eV) and an adiabatic migration barrier of 0.32~(eV) determined by nudged‑elastic‑band calculations. In contrast, we show that excess electrons do not self-trap on Nb-4$d$ orbitals, reflecting weak electron–phonon coupling in the conduction band manifold.
These results identify oxygen as an intrinsic hole trap in \ce{NaNbO3} and highlight the importance of including hole polarons in defect models of \ce{NaNbO3}-based electroceramics.
\end{abstract}

\date{\today}
\maketitle
\section{Introduction}
Increasingly stringent environmental regulations and the drive toward sustainable electronics have triggered an intensive search for lead-free
ferroelectric and antiferroelectric oxides for energy-storage and actuation applications~\cite{Shrout2007, Rodel2009, Bell_Deubzer_2018, Xu2017, YANG201972}. Among oxide antiferroelectrics, NaNbO$_3$ is particularly attractive because it is built from inexpensive, earth-abundant alkali elements~\cite{yang2020lead}. In addition, NaNbO$_3$ (NN) serves as the parent
phase of the (\text{K},\text{Na})NbO$_3$ piezoelectric family~\cite{Saito2004, Wu2015, Shirane, Koruza2020}, so processing routes and industrial know-how are already well established~\cite{fritsch2018electronic, liu2018antiferroelectrics, Li2013, Valdez}.

NN is a perovskite that can display ferroelectric, antiferroelectric, and paraelectric order depending on temperature and crystal symmetry~\cite{Megaw01011974, Lanfredi2002, Reznitchenko_2001, darlington1973low,glazer1973studies, glazer1972structure,koruza2010phase}. 
In its orthorhombic  phase~\cite{Luo2023, Johnston2010, Htet}, neighboring Nb‑centred dipoles are antiparallel, so the macroscopic polarization is zero. A strong electric field can align them parallel and create the ferroelectric phase. If the lattice relaxes back on field removal, the polarization–field trace forms a narrow double hysteresis loop. 
This property provides higher usable energy density and efficiency than the wide single loop of classical ferroelectric materials, especially for the energy storage of dielectric capacitors~\cite{liu2018antiferroelectrics}. The core problem is that once an electric field forces NN out of its antiferroelectric ground state, the induced ferroelectric phase becomes locked in, abolishing the reversible double‑loop hysteresis needed for efficient energy‑storage operation~\cite{Guo2015, Egert2020, Zhang2021}. This challenge arises because antiferroelectric and ferroelectric phases have comparable energy landscapes~\cite{shimizu2015lead}.
\begin{figure*} [!htp]
  \centering
  \subfigure[]{\includegraphics[width=.45\linewidth]{Figures/fig-1.pdf}\label{fig:Fig-1-a}}
  \subfigure[]{\includegraphics[width=.45\linewidth]{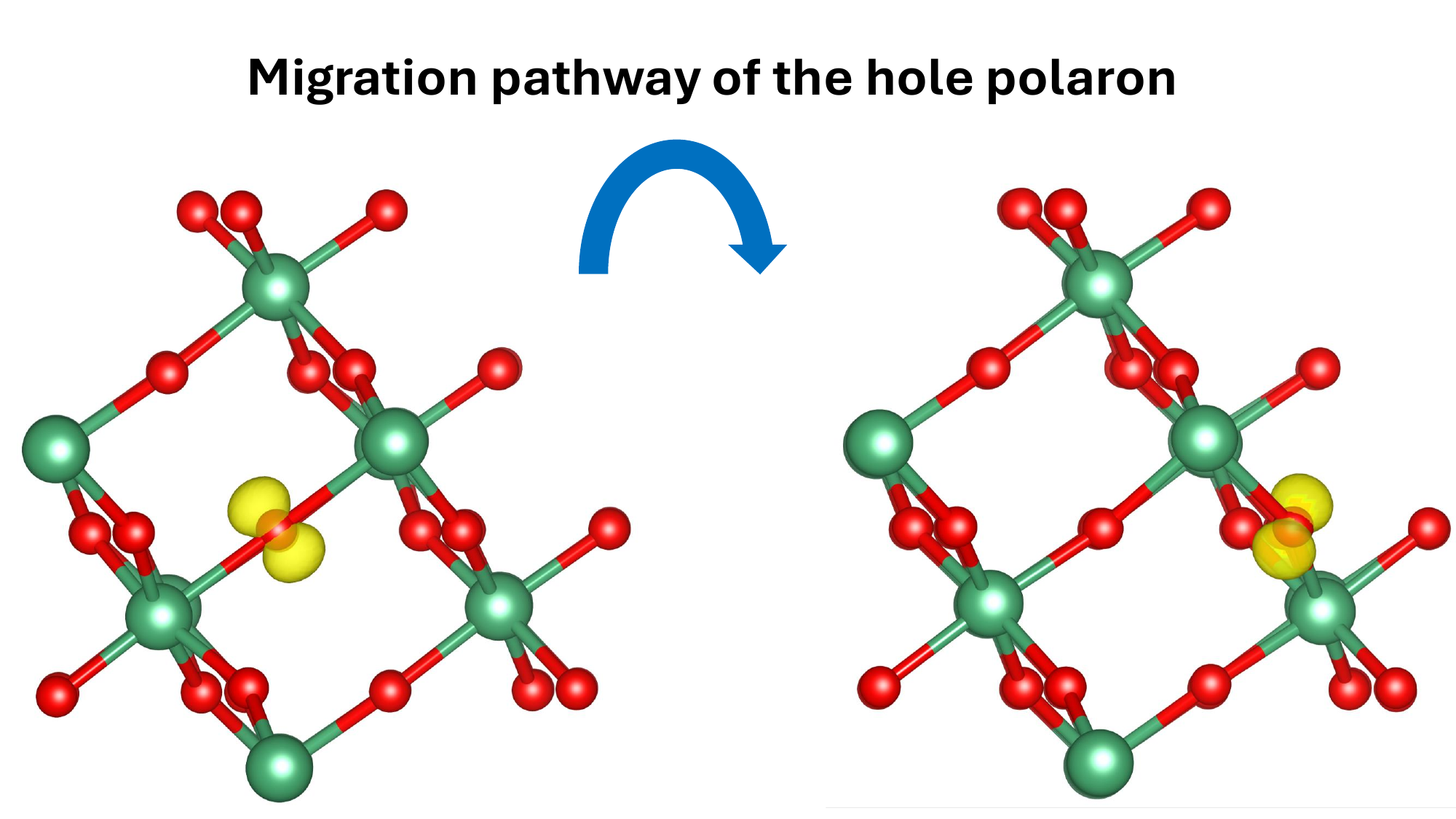}\label{fig:Fig-1-b}}
 \caption{
(a) Crystal structure of rhombohedral NaNbO$_3$ (space group \textit{R3c}) shown for a $3 \times 3 \times 3$ supercell. Green, blue, and red spheres denote Nb, Na, and O atoms, respectively. The localized hole polaron is represented by the yellow charge isosurface associated with O-$2p$ states. (b) Schematic illustration of the migration pathway of the hole polaron between neighboring oxygen sites.
}
  \label{Fig-1}
\end{figure*}

Among the possible solutions~\cite{Luo2023, Wang2025, KORUZA201777, Shiratori2005NaNbO3, shimizu2015lead, tan2018double}, one option is chemical composition modification by doping and co-doping~\cite{liu2018antiferroelectrics, Guo2015, Aso2023, Luo2023, Ye2019, Ye2025, qi2021emerging}. 
The idea is that in acceptor doped materials, frozen defect‑dipole complexes generate an internal field that returns the field‑induced ferroelectric phase to the antiferroelectric phase after bias removal, yielding a narrow, reversible double hysteresis loop~\cite{Luo2018,Villa2022}.
Therefore, by choosing appropriate dopants and co-dopants  and controlling charge compensation, one can turn NN into high-efficiency, lead‑free energy‑storage ceramics. 

In this context, a detailed understanding of the charge compensation mechanism is crucial because any imbalance in donor and acceptor charges provokes the formation of additional vacancies, raises leakage~\cite{SAKAMOTO20104256, Luo2018}, and ultimately erases the narrow double‑loop hysteresis~\cite{zhang2023tailoring}. While charge compensation is well established and routinely engineered in semiconductor materials, it has been far less explored in complex dielectric oxides such as NN.

Electronic charge compensation can occur through the self-trapping of excess electrons or holes by local lattice distortions, forming polarons~\cite{Devreese2015, Franchini2021, Dykman2015, alexandrov2010advances}. In this process, reduction or oxidation of a host ion is accompanied by a structural relaxation that lowers the carrier energy. The spatial extent of a polaron is governed by the balance between carrier delocalization and electron–phonon coupling~\cite{Emin_2012}. Strong local coupling favors small polarons, in which the excess charge and accompanying distortion are confined to approximately one lattice site~\cite{HOLSTEIN1959343}. Weaker coupling gives rise to large polarons, where the carrier remains extended over several unit cells~\cite{Fröhlich01031950}. Because a reliable first-principles description of large polarons requires substantially larger supercells, we restrict the present study to small-polaron formation.

Extensive experimental and first-principles work has clarified many structural, electronic, and defect-chemical aspects of NN, particularly for phases relevant near room temperature~\cite{ZHANG2020127, Villa2022, gouget2019isolating, shi2012photophysical, wang2016hybrid, BeinPRM2022}. By contrast, the intrinsic tendency of NN to form localized electronic carriers remains poorly understood. This is a significant gap because electron and hole polarons can act as charge-compensation species and may pin the Fermi level by defining the accessible range of electron and hole chemical potentials~\cite{Klein2023}. Such Fermi-level pinning can, in turn, affect conductivity, defect equilibria, and functional properties in oxide electroceramics~\cite{Hermans2020, Lohaus2018}. A quantitative assessment of carrier self-trapping energies is therefore required to establish whether polarons are relevant intrinsic charge carriers in NN.
Here, we address this question for the low-temperature rhombohedral phase of NN. Although the technologically relevant antiferroelectric response is associated with the orthorhombic phase, first-principles calculations for this structure are complicated by soft phonon modes and associated lattice instabilities~\cite{TADANO201918216, Pallikara_2022, Kotiuga2022}. The rhombohedral ground-state phase provides a structurally well-defined reference system and therefore offers a controlled starting point for assessing small-polaron formation in NN.

A further incentive to explore polaron study in rhombohedral NN comes from its sister compound \ce{LiNbO3}, where the full hierarchy of polaron states has already been mapped out~\cite{Schirmer2009, Schmidt2020, Sanna2015, Sanna20155, Krampf_2021, Sanna2017}. 
Optical-absorption and EPR measurements, supported by first-principles calculations, identify three distinct small electron polaron species in \ce{LiNbO3}. These are free small polarons that self-trap at regular Nb sites, bound small polarons localized at antisite \ce{Nb_{Li}} defects, and bipolarons in which an electron pair is shared between an \ce{Nb_{Li}} defect and a neighboring Nb ion.
These fingerprints of polarons govern photorefractive response and multi-color holographic recording, demonstrating how point‑defect chemistry dictates macroscopic electro‑optic performance in niobates.
Because rhombohedral NN shares the same Nb–O octahedral framework, a systematic search for analogous free, bound and defect‑bound polarons promises both fresh fundamental insight and transferable defect‑engineering strategies for NN as lead‑free capacitors.

In this study, we employ DFT+$U$ calculations to investigate the formation and stability of small electron and hole polarons in NN in its low-temperature rhombohedral phase.  Our results show that the small hole polaron is stable in rhombohedral NN, localizing on O $2p$ orbitals with a trapping energy of $-0.65$ (eV) and migrating with an activation barrier of 0.32 (eV). In contrast, there is no small electron polaron located  on Nb $4d$ orbitals, revealing a strong asymmetry in charge localization that significantly influences the material’s electronic and transport behavior.
The paper is structured as follows. Sec.~\ref{section_II} presents the computational methods, including the details of the DFT+$U$ calculations. Sec.~\ref{section_III} discusses the formation, stability, and migration of small hole polarons, along with the electronic structure and charge localization behavior. 
Also, in this section, we show why the small electron-polaron on Nb is not stable. 
Finally, Sec.~\ref{section_IV} provides a summary of the main findings.
\section{Computational Details}
\label{section_II}
\begin{figure*} [!htp]
  \centering
  \subfigure[Without finite-size correction]{\includegraphics[width=.45\linewidth]{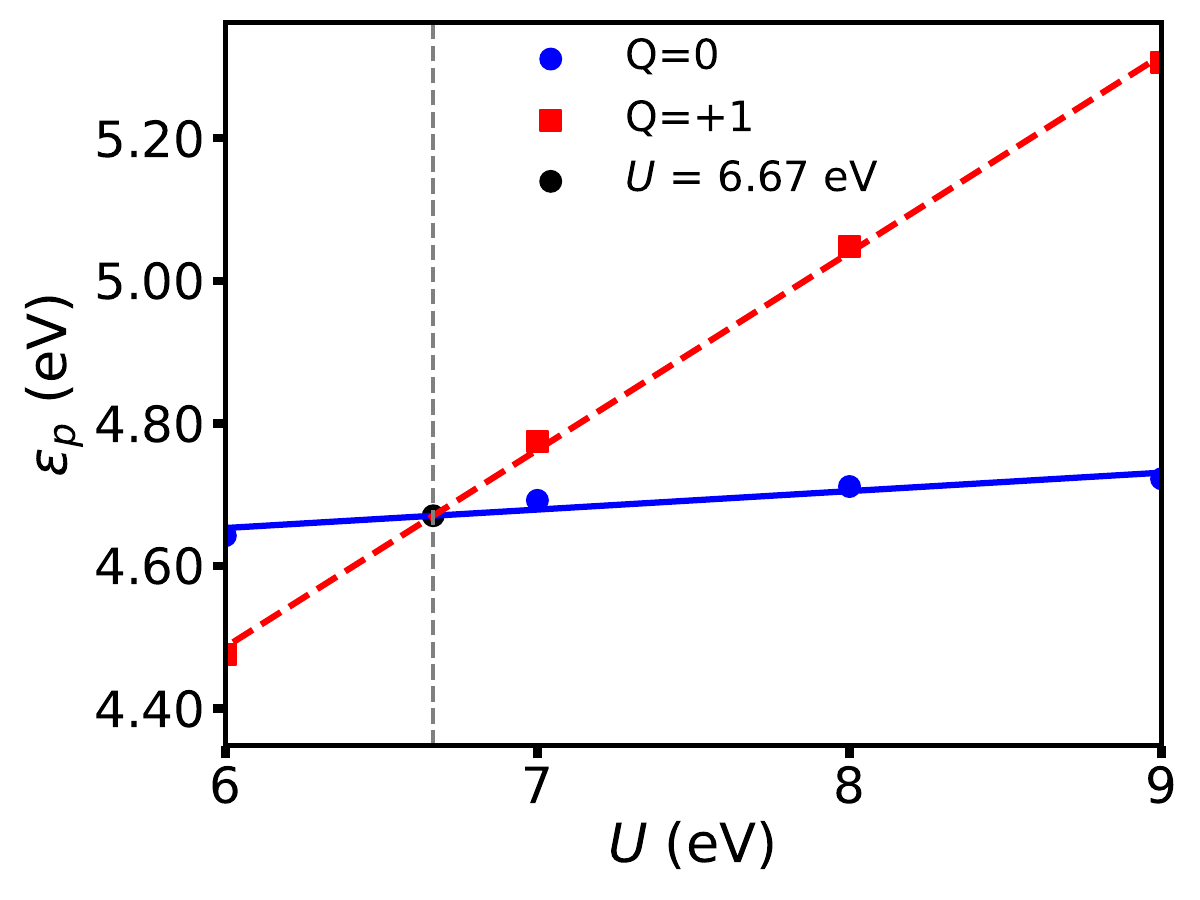}\label{fig:Fig-2-a}}
  \subfigure[With finite-size correction]{\includegraphics[width=.45\linewidth]{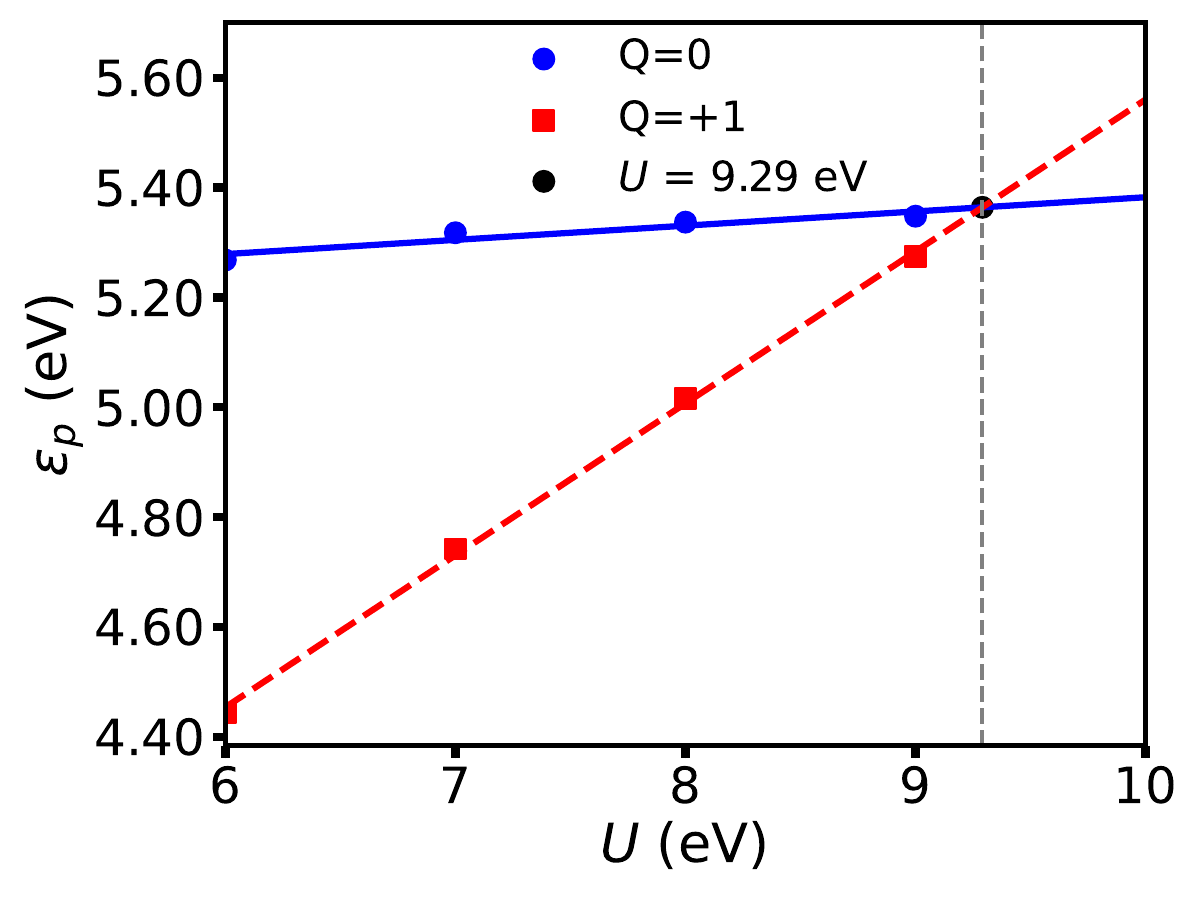}\label{fig:Fig-2-b}}
 \caption{
Determination of the optimal $U$ for a small hole polaron in rhombohedral NN. Panels (a) and (b) illustrate the piecewise-linearity condition without and with finite-size corrections, respectively, resulting in an optimal value of $U=9.29$ (eV) of O-$2p$.}
  \label{Fig-2}
\end{figure*}
All first-principles calculations are performed within the framework of density functional theory (DFT) as implemented in the Vienna \textit{ab initio} simulation package (VASP)~\cite{KRESSE199615}. The interaction between ions and electrons is described using the projector augmented-wave (PAW) approach~\cite{Bloch_1994, Kresse_1999}.
The exchange-correlation functional is treated using the generalized gradient approximation (GGA)~\cite{Perdew_1996} within the Perdew--Burke--Ernzerhof revised for solids (PBEsol) parametrization~\cite{Perdew2008}.
The pseudopotentials include valence electron configurations corresponding to $Z_{\text{val}} = 6$ for O, $Z_{\text{val}} = 1$ for Na, and $Z_{\text{val}} = 13$ for Nb.

Polaronic states are modeled by adding or removing a single electron from the supercell and initializing a localized magnetic moment on a specific atomic site to promote charge localization. 
To this end, we follow the bond distortion method~\cite{Pham2020}, whereby selected bond lengths are intentionally elongated to create a local potential well that favors charge localization. For the small hole polaron  on O, one electron is removed from the supercell, and initial distortions are introduced by increasing both Na–O and Nb–O bond lengths around a specific O site. For the small electron polaron on Nb, one electron is added to the system, and the surrounding Nb–O bonds of a selected Nb site are elongated to facilitate electron trapping. These structural distortions, combined with charge localization, assist in stabilizing localized polaronic states during ionic relaxation. 

The localization of excess charge carriers is not obtained within semilocal GGA, which favors delocalized solutions even when local lattice distortions are imposed. We therefore employ the DFT+$U$ approach to correct the self-interaction error and enhance the description of localized Nb-$4d$ and O-$2p$ states relevant for electron and hole polarons, respectively. The Hubbard correction is applied within the rotationally invariant Dudarev formalism, using a bare $U$ parameter~\cite{Anisimov_1991, Dudarev_1998}.

To initialize localized polaronic configurations, we control the occupation matrix following the approach of Allen and Watson~\cite{allen2014occupation}. In a first step, the ionic positions are relaxed under constrained orbital occupations that enforce localization of the excess carrier on a selected Nb or O site. After the convergence of this constrained calculation, the occupation-matrix constraint is removed, and the structure is relaxed again. A polaron is considered stable only if the localized charge distribution and associated lattice distortion persist after this unconstrained relaxation.

Choosing a $U$ parameter that correctly captures electron–electron correlations in a localized polaronic state is non-trivial. Tuning $U$ to reproduce bulk observables such as the band gap does not guaranty an accurate description of the defect orbital itself.  To overcome this problem, we adopt the enforced piecewise-linearity protocol proposed by Falletta \textit{et al.} for polaronic defect states~\cite{Falleta_2020, Falletta2022}.  In this approach, the optimal $U$ is the one that restores the piecewise linearity condition of the total energy with respect to the occupation of the polaron level once all spurious finite-size interactions have been removed from both the charged and neutral supercells~\cite{Falleta_2020}. 
Since this approach is based on high-frequency and static dielectric constants, we employ density functional perturbation theory (DFPT) within VASP to determine these constants. The ionic contribution to the static dielectric tensor is obtained by evaluating the force‑constant matrices and the internal‑strain tensors via density-functional perturbation theory.

All structures are fully relaxed until the forces on each atom are less than $0.01$~(eV/\AA). 
Geometry optimization of the primitive cell (including 10-atoms) is performed using a higher precision, employing a plane-wave energy cutoff of 900 (eV) and a $10 \times 10 \times 10$ $k$-point mesh.
For the investigation of the stability of small hole and electron polarons, we use a $3 \times 3 \times 3$ supercell, consisting of 270 atoms, with a plane-wave energy cutoff of 500~(eV) and a $\Gamma$-centered $2 \times 2 \times 2$ $k$-point mesh. 

After installing the polaron, we determine minimum-energy pathways and activation barriers using the nudged elastic-band (NEB) method~\cite{jonsson1998nudged, mills1995reversible} as implemented in VASP code. We generate five intermediate images by linear interpolation between fully relaxed initial and final structures, resulting in a band of seven images in total. 
During the band optimization, we use a variable spring constant (SPRING = –5 (eV/\AA$^{2}$)) and activate the climbing-image procedure to accurately locate the saddle point.
We converge the electronic self-consistency for each image to $10^{-6}$ (eV) and relax the atomic positions until the residual forces perpendicular to the band are smaller than 0.01(eV/\AA). All other computational parameters (exchange–correlation functional, $U$ parameter, plane-wave cutoff, $k$-point mesh, and PAW potentials) are kept consistent with those described earlier.
\section{Results and discussion}
\label{section_III}

In the following, we first establish the structural and dielectric properties of rhombohedral NaNbO$_3$, which provide the reference system and screening parameters required for the finite-size correction of charged polaronic supercells. We then analyze hole self-trapping on oxygen sites, including the determination of the Hubbard parameter from enforced piecewise linearity, the resulting trapping energy, electronic structure, and migration barrier. Finally, we examine the possibility of electron self-trapping on Nb sites and relate its instability to the dispersion of the Nb-derived conduction-band states.

\subsection{Structural relaxation and dielectric response}
\begin{figure*}[!htp]
  \centering
  \subfigure[Delocalized charge]{%
    \includegraphics[width=.45\linewidth]{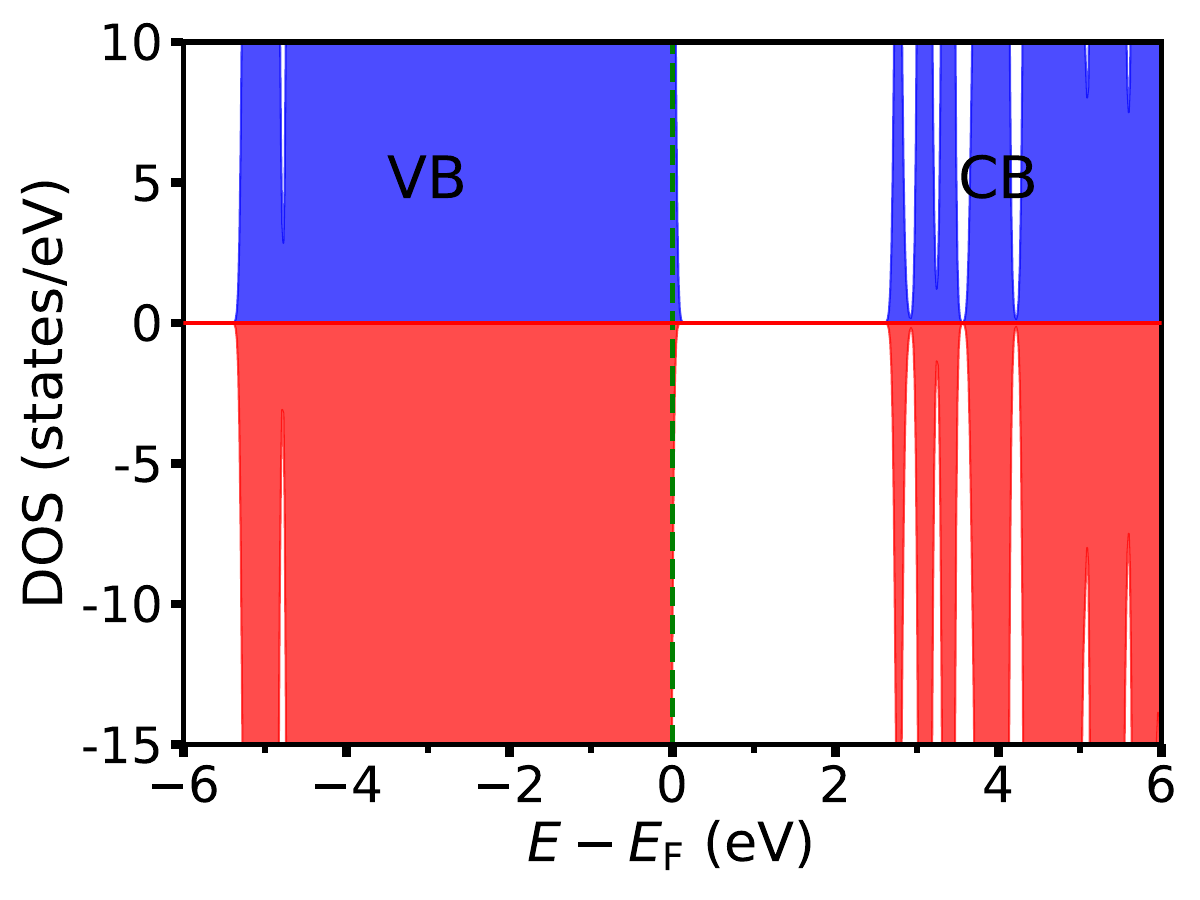}%
    \label{fig:Fig-3-a}
  }
  \subfigure[Localized charge]{%
    \includegraphics[width=.45\linewidth]{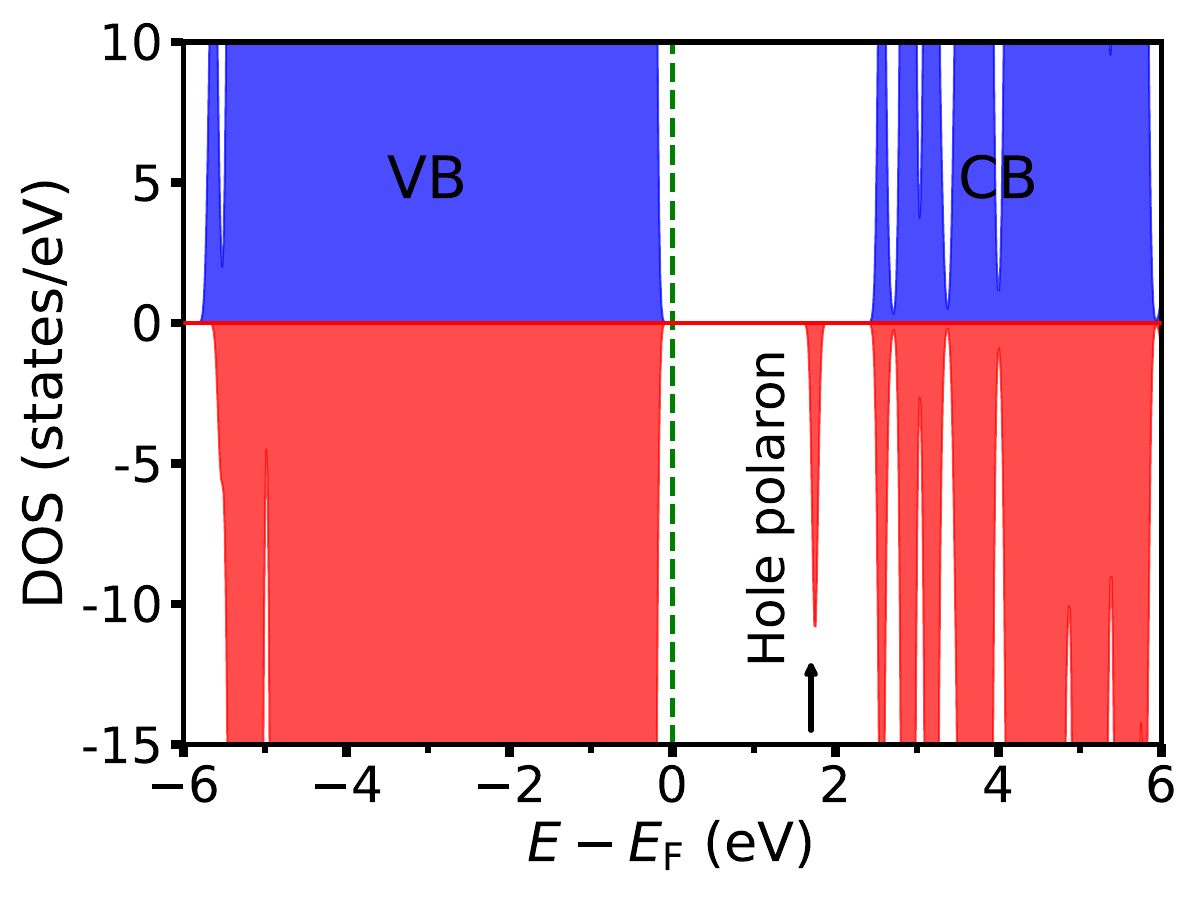}%
    \label{fig:Fig-3-b}
  }
  \caption{
    Comparative density of states (DOS) for the (a) delocalized and (b) localized hole-polaron states. Spin-up and spin-down channels are shown in blue and red, respectively, and the Fermi level is set to 0~(eV), indicated by the green dashed line. In the delocalized configuration, the valence band crosses the Fermi level, consistent with metallic behavior arising from a hole distributed uniformly over the lattice. By contrast, in the localized small-polaron state, a distinct hole-polaron level appears 1.95~(eV) above the valence-band maximum.
  }
  \label{Fig-3}
\end{figure*}

We use the low-temperature rhombohedral phase of NN (space group R3c) determined by neutron diffraction at $T = 12$ K by Mishra \textit{et al.}~\cite{Mishra_2007} as the starting structure for all first-principles calculations. Lattice vectors and internal coordinates are fully relaxed in the 10-atom primitive cell shown in Fig.~\ref{Fig-1} using the PBEsol exchange–correlation functional. The optimized lattice parameters differ from experiment by less than 0.4\% and retain the R3c symmetry (Tab.~\ref{tab:lattice}). In this phase NN is ferroelectric, with polar displacements of Na, Nb,
and O along the pseudocubic [111] direction\cite{Mishra_2007}.

\begin{table}[!htp]
\centering
\caption{Optimized lattice parameters of rhombohedral NN.  Experimental values from Ref.~\cite{Mishra_2007} are given in parentheses.}
\begin{tabular}{@{}ll@{}}
\toprule
\text{Lattice constants (\AA)} & \text{Angles ($^{\circ}$)} \\ \midrule
$a = 5.463$ \;(5.481) & $\alpha = 60.547$ \;(60.421) \\
$b = 5.463$ \;(5.481) & $\beta = 60.547$ \;(60.421) \\
$c = 5.548$ \;(5.552) & $\gamma = 59.972$ \;(60.000) \\ \bottomrule
\end{tabular}
\label{tab:lattice}
\end{table}

The static dielectric tensor has been computed using DFPT within the DFT+$U$ framework, with a Hubbard correction of $U = 4.0$~(eV) applied to the Nb $4d$ states; this choice is consistent with prior DFT+$U$ studies of the closely related sister compound LiNbO$_3$, in which $U = 4.0$~(eV) has likewise been adopted for Nb $4d$ states~\cite{Krampf_2021, Nahm2008,  You2019}. Before the DFPT step, only the internal coordinates are re-relaxed at the DFT+$U$ level while the lattice constants are fixed to the PBEsol values. The electronic contribution $\varepsilon_{\infty}$, the lattice contribution $\varepsilon_{\text{ion}}$, and their sum $\varepsilon$ are listed in Tab.~\ref{tab:dielectric}.
For the R3c phase, the magnitude of the electronic dielectric constant
inferred from the low-frequency limit of the optical dielectric function
calculated by Fritsch \textit{et al.}\cite{Fisher2018} is of the order
$5–6$, in good agreement with our computed value of
$\varepsilon_{\infty} = 5.28$–$5.4$.

\begin{table}[!htp]
\centering
\caption{Electronic ($\varepsilon_{\infty}$), ionic ($\varepsilon_{\text{ion}}$), and static ($\varepsilon$) dielectric constants of rhombohedral NN obtained from DFT+$U$ calculations.}
\begin{tabular}{@{}cccc@{}}
\toprule
\text{Direction} & $\boldsymbol{\varepsilon_{\infty}}$ & $\boldsymbol{\varepsilon_{\text{ion}}}$ & $\boldsymbol{\varepsilon}$ \\ \midrule
$a$ & 5.28 & 51.80 & 57.08 \\
$b$ & 5.28 & 51.39 & 56.67 \\
$c$ & 5.40 & 111.26 & 116.66 \\ \bottomrule
\end{tabular}
\label{tab:dielectric}
\end{table}
\subsection{Small hole polaron on O}
In pure single-crystal rhombohedral NN, oxygen ions are nominally in the
$\mathrm{O}^{2-}$ oxidation state. Formation of a small hole polaron corresponds
to local oxidation toward an $\mathrm{O}^{-}$--configuration, where the hole
is predominantly localized in one of the O-$2p$ orbitals.
Standard GGA functionals often favor an artificially delocalized hole due to
self-interaction error and the insufficient description of on-site electronic
correlations in charged cell. Therefore, an explicit correction
is required to stabilize the localized polaron.
Here, we adopt the DFT+$U$ framework as a computationally efficient scheme to
capture the required localization. We find that polaron configurations
initialized in the $p_x$, $p_y$, and $p_z$ orbitals are nearly degenerate, with
total-energy differences on the order of $10^{-4}\,\mathrm{eV}$. Given this
near-degeneracy, we select the $p_x$-localized hole polaron as a representative
configuration, and all subsequent calculations are performed using this $p_x$
polaron state.
\subsubsection{Determination of the $U$ parameter via enforced piecewise-linearity}

For an exact functional, the combination of piecewise linearity and Janak's
theorem~\cite{Janak_1978} implies that the energy of the localized (polaron) level \(\epsilon_p(q)\) is
constant between integer occupations. We therefore enforce
\(\epsilon_p(Q)=\epsilon_p(0)\), with \(Q=+1\) (\(-1\)) for a hole (electron)
polaron as shown in Fig.~\ref{Fig-2}. 
Here, we focus on hole-polaron formation on O-$2p$ states and determine the corrected $U$ parameter for these orbitals in order to evaluate the trapping energy.
After applying finite-size correction on occupied and unoccupied polaron level, the optimal $U$ parameter on O-$2p$ increases from the uncorrected value of $U=6.67$ (eV) to the corrected value  of 9.29 (eV), underscoring the importance of properly treating electrostatic finite-size effects.
All subsequent polaron calculations, including the trapping energy reported below,
employ the non-empirically derived value
$U_{\mathrm{O}-2p}=9.29~\mathrm{(eV)}$.
This value is comparable to the $U$ values used for O-2$p$ hole polarons in other oxides~\cite {Deskins, Paul2014}, especially \ch{LiNbO3}~\cite{cryst12111586}.

\begin{figure} [!htp]
  \centering
  \includegraphics[width=0.94\linewidth]{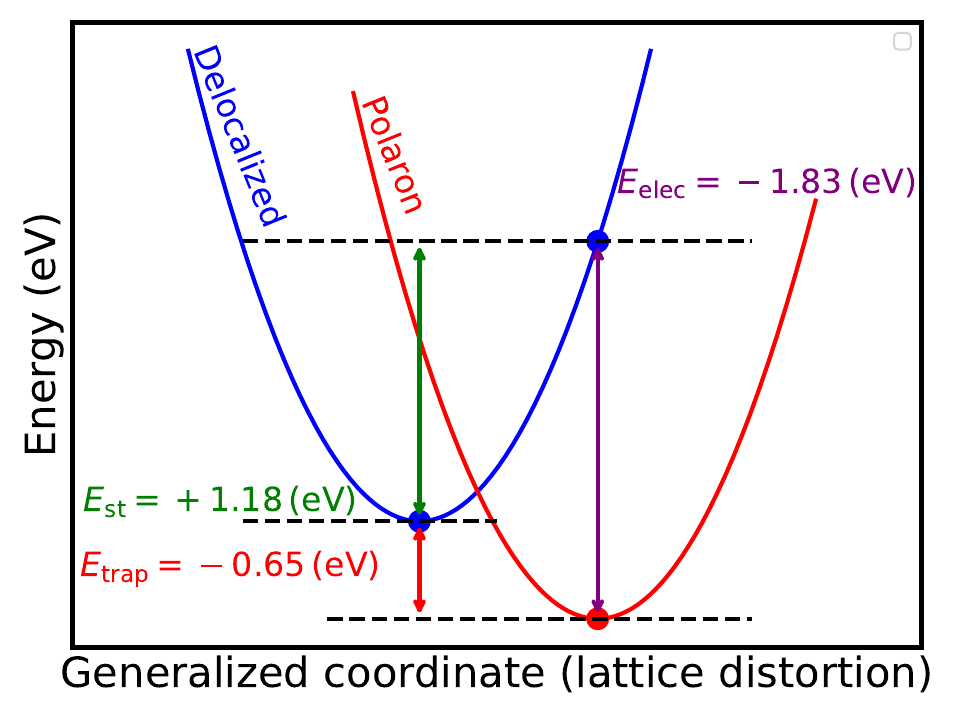}
  \caption{Configuration-coordinate diagram illustrating the decomposition of the trapping energy into its electronic and lattice strain contributions.}    
  \label{fig:Fig-4}
\end{figure}
\subsubsection{Trapping energy and stability of small hole polaron}
Fig.~\ref{fig:Fig-3-a} and Fig.~\ref{fig:Fig-3-b} display the density of states (DOS) for the localized and delocalized configurations, respectively.  
The delocalized state corresponds to an extra hole being uniformly distributed across all ions, resulting in a metallic state where the Fermi energy crosses the valence band.  
In contrast, when the extra hole is localized on a specific O atom, forming a small hole-polaron, a distinct state appears within the band gap, and the Fermi energy does neither cross the valence band nor the conduction band (Fig.~\ref{fig:Fig-3-b}).
For a stable small hole-polaron, the total energy of the localized configuration must be lower than that of the delocalized one.

\begin{figure} [!htp]
  \centering
  \includegraphics[width=0.95\linewidth]{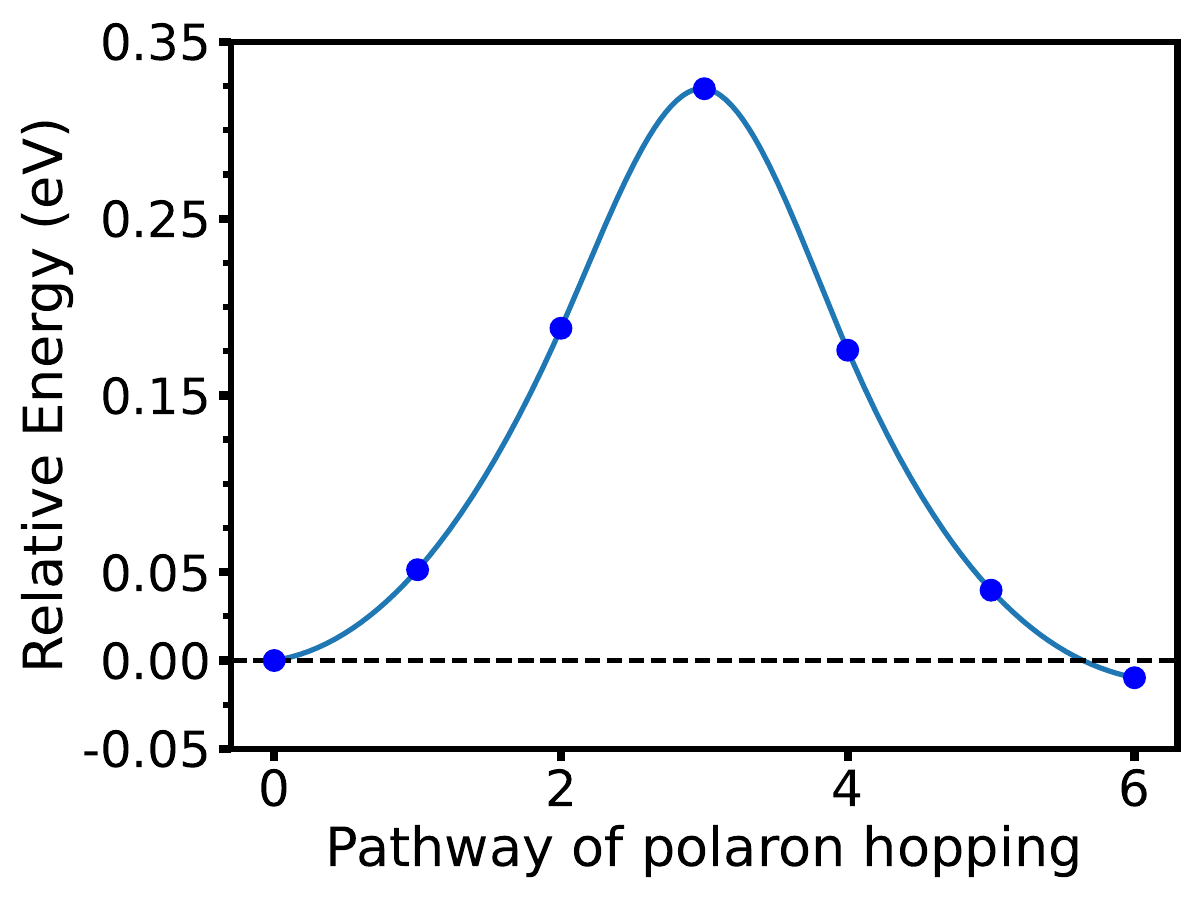}
  \caption{The plot shows the relative energies from the NEB calculation for the migration of a small hole polaron between two nearest-neighbor, symmetry-equivalent O sites. Here, 0 on the x-axis corresponds to the localized hole polaron at one O site (O‑2$p$), while 1 corresponds to the localized hole polaron at the nearest-neighbor site.}    
  \label{fig:Fig-5}
\end{figure}

\begin{figure*} [!htp]
  \centering
  \subfigure[\ce{NaNbO3}]{\includegraphics[width=.45\linewidth]{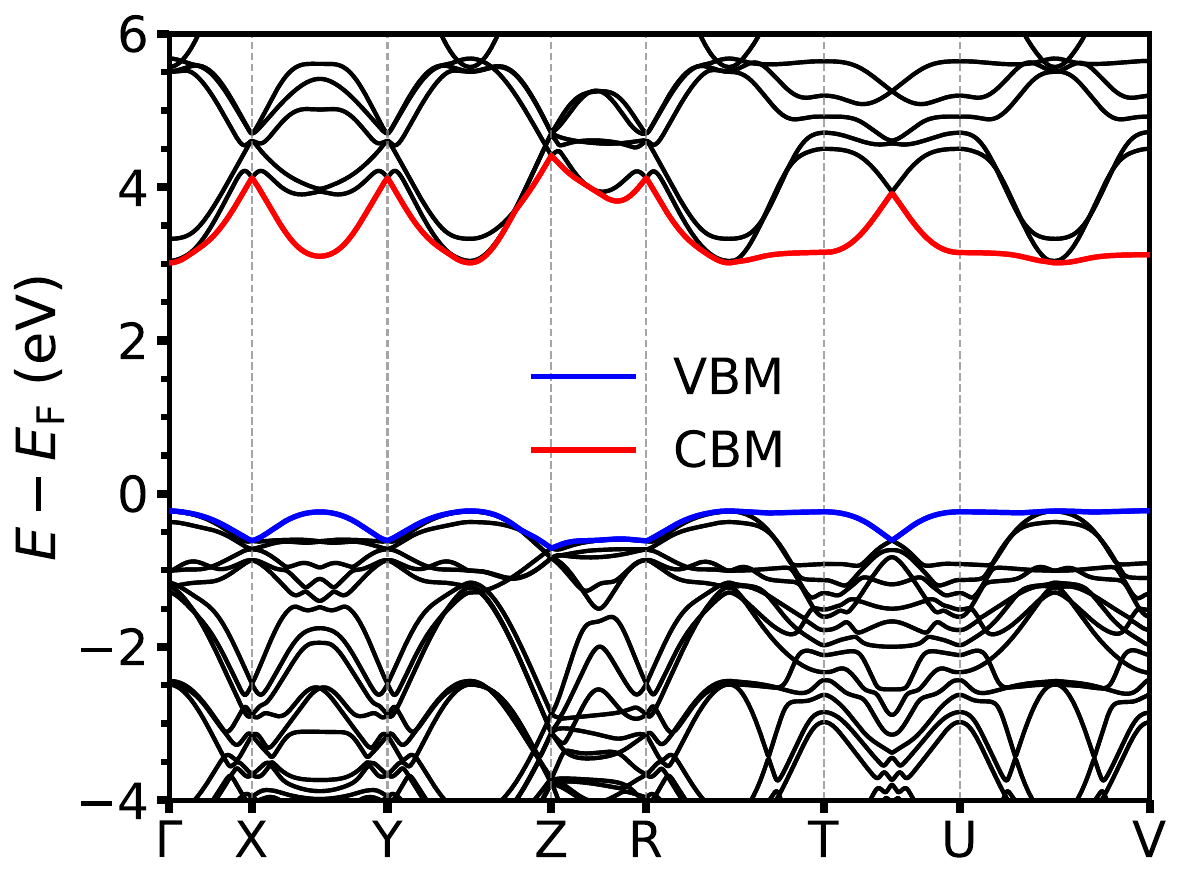}\label{fig:Fig-6-a}}
  \subfigure[\ce{LiNbO3}]{\includegraphics[width=.45\linewidth]{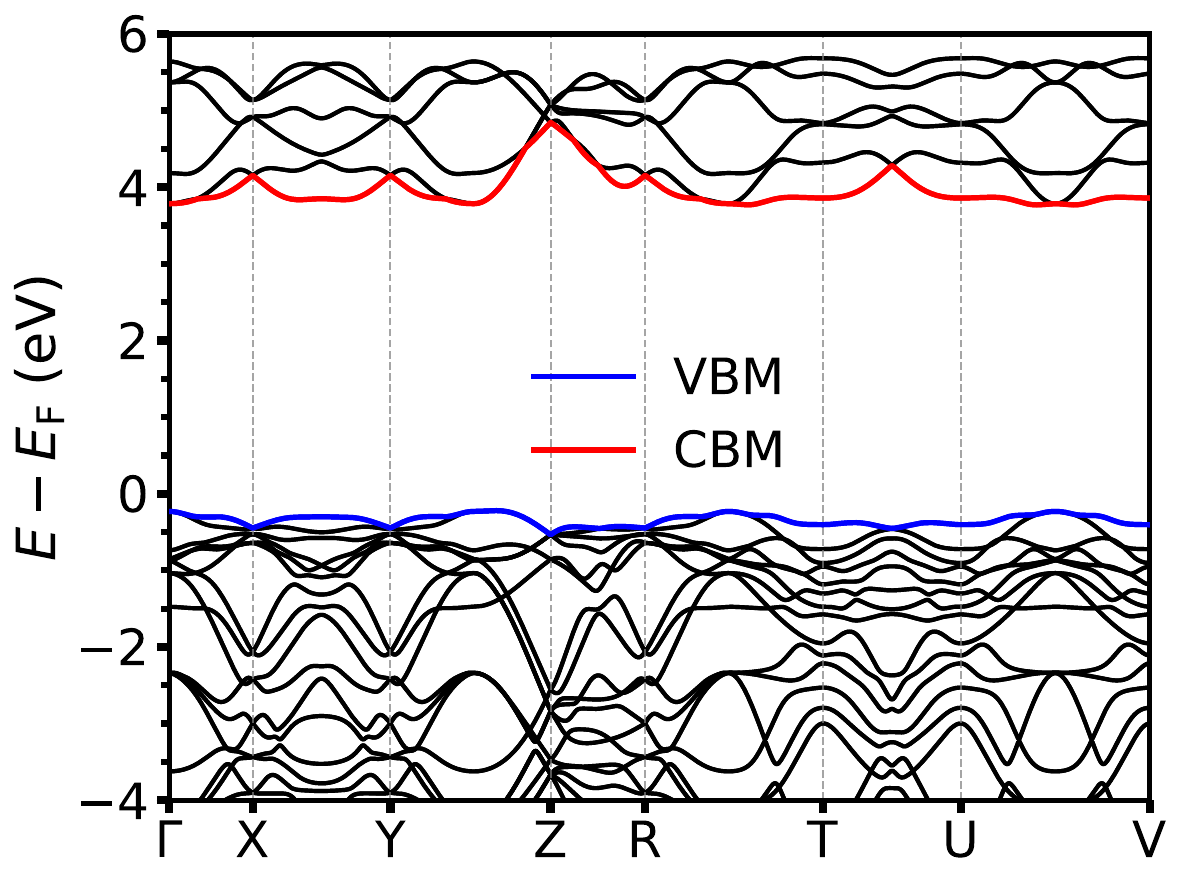}\label{fig:Fig-6-b}}
 \caption{ The electronic DFT$+U$ band structures for the rhombohedral
R3c phase of (a)~NaNbO$_3$ and (b)~LiNbO$_3$.
The VBM and CBM are highlighted.
LiNbO$_3$ exhibits a noticeably flatter Nb‑4$d$ CBM
than NaNbO$_3$,
yielding a heavier electron effective mass and stronger
electron–phonon coupling.
The high‑symmetry k‑point path is generated with the SeeK‑path web interface\cite{seekpath}.
}
  \label{Fig-6}
\end{figure*}
To determine whether this type of polaron is stable, we calculate the trapping energy \( E_{\rm trap} \).  
Obtaining this number from the difference between the
polaronic configuration and the delocalized hole at the band edge would require
treating the latter as a metallic state, with a partially depleted valence
band. Such metallic occupations converge very slowly with respect to
$k$-point sampling in supercell calculations and are therefore numerically
inefficient~\cite{Paul2014}. We  instead
obtain the hole–polaron trapping energy $E_{\mathrm{trap}}$ as in Refs.~\cite{Osterbacka2020, Paul2014} from the following relation,
\begin{align}
E_{\mathrm{trap}}
  &= E_{\mathrm{pol}}^{+1} - E_{\mathrm{pristine}}^{0}
     + \epsilon_{\mathrm{VBM}} + E_{\mathrm{correction}},
\label{eq:trapping}
\end{align}
where $E_{\mathrm{pol}}^{+1}$ and $E_{\mathrm{pristine}}^{0}$ are the total
energies of supercells containing, respectively, a localized (polaronic) hole
in the $+1$ charge state and the pristine neutral system, and
$\epsilon_{\mathrm{VBM}}$ is the valence-band-maximum eigenvalue of the
pristine cell.
$E_{\mathrm{correction}}$   is the electrostatic finite-size correction, which has been computed following Ref.~\cite{Falleta_2020} and equals 
+0.02 (eV).
Using Eq.~(\ref{eq:trapping}), we obtain a trapping energy of
$E_{\mathrm{trap}} = -0.65$~(eV) for the small hole polaron, indicating that the localized configuration is energetically favored over the delocalized
hole at the valence band.

Fig.~\ref{fig:Fig-4} shows the generalized coordinate diagram illustrating the different energy components. The trapping energy can be decomposed into a strain energy (\(  E_{\rm st} \)) and an electronic energy (\(  E_{\rm elec} \)).  
\(E_{\rm st} \) is calculated by comparing the total energies of the polaronic configuration without extra charge and the ideal structure without extra charge.  
This term reflects the energy gain associated with the lattice distortion required to trap the extra charge.  
Our calculations show that \( E_{\rm st} = +1.18\,\mathrm{(eV)} \).
The electronic energy \(\ E_{\rm elec} \) is a negative quantity representing the energy required to excite the trapped hole from its localized state to the valence band.  
Our calculations show that \( E_{\rm elec} = -1.83\,\mathrm{(eV)} \).  

While the configuration-coordinate analysis confirms that hole self‑trapping is energetically favorable, the practical impact on charge transport depends on how swiftly these polarons can hop between lattice sites. We therefore quantified the minimum‑energy migration pathway—and its activation barrier—using the NEB method.
Fig.~\ref{fig:Fig-5} displays the relative energies obtained from the NEB calculation for the migration of a small hole-polaron between two nearest-neighbor symmetry-equivalent O sites in rhombohedral NN. The energy increases monotonically from the initial configuration (image 0) to a maximum of 0.32 (eV) at the saddle point (image 3), after which it decreases symmetrically toward the final configuration (image 6). Owing to the equivalence of the initial and final states, the energy profile is nearly symmetric, and the forward and backward migration barriers are identical.
The calculated activation energy of  0.32~(eV) suggests that thermally activated hopping of hole polarons becomes possible at temperatures above room temperature. 
\subsection{Electron polaron on Nb}
Having established that holes can self-trap on oxygen sites, we next examine the opposite charge state: localization of an excess electron on Nb. In nominally ionic NaNbO$_3$, this process corresponds to the local reduction of Nb$^{5+}$ toward an Nb$^{4+}$-like configuration, with the additional electron occupying a 4$d^1$ state. This implies that the excess electron would occupy one of the available Nb 4$d$ sublevels (i.e., $d_{xz}$, $d_{yz}$, $d_{xy}$, $d_{x^2 - y^2}$, or $d_{z^2}$).
Whether such a state is stable depends on the balance between the electronic energy gained by localization and the elastic energy required to distort the surrounding NbO$_6$ octahedron. We therefore test the stability of Nb-centered electron polarons by initializing localized occupations in the Nb-$4d$ state and subsequently relaxing the structure after removal of the occupation constraint.

For this, we consider a $3 \times 3 \times 3$ supercell explicitly incorporating Nb–O bond distortions, as described in Sec.~\ref{section_II}.  To investigate the stability of this type of polaron, we constrain occupation matrices for each distinct type of Nb-4$d$ orbital during DFT+$U$ calculations.  After the constrained calculations have converged, we remove these constraints and reoptimize the geometry to check whether the localized charge remains stable. 
Our results show that, for all orbital occupations, the extra electron does not remain localized once the constraints are removed.   This finding indicates that an electron polaron on Nb in rhombohedral NN is not stable.

This instability arises primarily because of electron–phonon coupling
in Nb-4$d$ orbitals, the coupling is weaker compared to hole–phonon coupling in O-2$p$ orbitals. 
In polar crystals, the strength of long-range electron–phonon coupling can be  characterized by the dimensionless Fröhlich coupling constant~\cite{Frohlich01071954,Emin_2012,Kokott_2018},
\begin{align} 
\alpha=\frac{e^2}{\hbar}(\frac{1}{\epsilon_{0}}-\frac{1}{\epsilon_{\infty}})\sqrt{\frac{m^{band}}{2\hbar w}}
\label{Equation-1} 
\end{align}
where, $m^\mathrm{band}$ is band effective mass and $\hbar$, $w$ and $e$ are reduced Planck constant,  angular frequency of the longitudinal optical phonon mode and elementary charge of the carrier, respectively.
Because $\alpha$ increases with the square root of the carrier band effective
mass $m^\text{band}$, heavier bands experience stronger electron–phonon coupling~\cite{Emin_2012}.
In the weak-coupling regime, the carrier forms a large
polaron, while for larger $\alpha$, the coupling becomes strong enough to favor self-trapping and the formation of a small polaron ~\cite{Kokott_2018}.

Fig.~\ref{fig:Fig-6-a} presents the band structure of the
rhombohedral NN primitive cell.
According to Fig.~\ref{fig:Fig-6-a}, the valence-band maximum (VBM, blue) is nearly dispersionless along the entire path, which indicates a very large hole effective mass and, consequently, strong coupling to local lattice distortions. In contrast, the conduction-band minimum (CBM, red) exhibits a sizeable dispersion, consistent with a substantially lighter electron effective mass and much weaker electron–phonon coupling. The VBM is dominated by O-$2p$ states, whereas the CBM originates mainly from Nb-$4d$ orbitals. This difference in band curvature is therefore consistent with our DFT+$U$ results for charged supercells, which show that holes readily self-trap on O and form stable small polarons, while excess electrons do not localize on Nb and electron polarons remain unstable even when localized occupation matrices are imposed as initial conditions.

To elucidate the role of the Nb-4$d$ conduction-band dispersion in electron-polaron formation, we also examine LiNbO$_3$, a prototypical electron-polaron host with the same R3c symmetry~\cite{Krampf_2021}. Its calculated band structure (Fig.~\ref{fig:Fig-6-b}) shows that the bandwidth of the conduction-band minimum is smaller than in NaNbO$_3$, indicating markedly flatter Nb-4$d$-derived states. The associated heavier electron effective mass enhances the electron–phonon coupling, so that the excess electron localizes on a Nb site and reduces Nb$^{5+}$ to Nb$^{4+}$, thereby forming a small electron polaron.
\section{Conclusion}
\label{section_IV}
This work reveals a carrier-selective tendency toward self-trapping in rhombohedral NaNbO$_3$. Using a finite-size-corrected, piecewise-linearity-based DFT+$U$ scheme, we obtain a nonempirical value of $U=9.29$ (eV) for the O-$2p$ states, enabling a reliable description of the localized oxygen-hole state. The resulting hole polaron is energetically stable, with a trapping energy of $-0.65$ (eV), an in-gap level located 1.95 (eV) above the valence-band maximum, and a migration barrier of 0.32 (eV). These results indicate that holes in NaNbO$_3$ are not purely band-like carriers, but can self-trap as mobile, thermally activated oxygen-centered small polarons.

In contrast, excess electrons do not form stable Nb-centered small polarons once occupation constraints are removed, even when local Nb--O distortions are imposed. 
This instability is attributed to the relatively dispersive Nb-derived conduction-band minimum of NaNbO$_3$, which gives rise to a light electron effective mass and weak electron--phonon coupling. Consequently, lattice relaxation around a Nb site is insufficient to compensate the energetic cost of electron localization. Hence, unlike in related niobates such as \ch{LiNbO3},
free Nb-centered electron polarons are not stable in rhombohedral \ch{NaNbO3}. But small hole polarons behave differently because the O-derived valence-band maximum is considerably flatter, leading to stronger electron--phonon coupling and stable self-trapping on oxygen.

These findings establish a microscopic picture of electronic charge localization in \ch{NaNbO3}, which is important for defect models of \ch{NaNbO3}-based electroceramics, where the balance between ionic defects, dopants, and localized electronic carriers controls leakage, internal bias fields, and ultimately functional performance.
The absence of an intrinsically stable  Nb-centered small electron polaron does not rule out other electron-localization mechanisms in NaNbO$_3$. Excess electrons may instead be stabilized as large polarons or as defect-bound Nb-centered polarons in the presence of vacancies, dopants, or other extrinsic trapping centers.

\section*{Acknowledgement}
MA, JR, and KA acknowledge financial support from the Collaborative Research Center FLAIR (Fermi level engineering applied to oxide electroceramics), funded by the German Research Foundation (DFG) under Project-ID No. 463184206–SFB 1548 (project A02).
Lichtenberg Supercomputing Center is gratefully acknowledged
by MA as the provider of needed computing facilities.
MA sincerely thanks Delwin Perera for his discussions and suggestions.

\bibliographystyle{apsrev4-1}
\bibliography{bib.bib} 
\end{document}